\documentstyle[preprint,aps]{revtex}
\begin{document}
\title{Dependence of the Shell-Model Single-Particle Energies on Different 
Components of the Nucleon-Nucleon Interaction}
\author{Y.Y Sharon$^1$, M.S. Fayache$^2$, Y. D. Devi$^1$, L. Zamick$^1$ 
and H. M\"{u}ther$^3$\\
(1) Department of Physics and Astronomy, Rutgers University
Piscataway, New Jersey 08855\\
\noindent (2) D\'{e}partement de Physique, Facult\'{e} des Sciences de Tunis\\
Tunis 1060, Tunisia\\ \noindent (3) Institut f\"{u}r Theoretische Physik 
der Universit\"{a}t T\"{u}bingen, D-72076 T\"{u}bingen, Germany}
\date{\today}
\maketitle
\begin{abstract}
The spin-orbit splittings in the spectra of nuclei with mass numbers 5, 15 and 
17 are studied within the framework of shell-model configuration mixing 
calculations including 2$\hbar \omega$ excitations. 
The contributions of the two-body spin-orbit and tensor components of the 
nucleon-nucleon 
interaction are studied in various model spaces. It is found that the effects
of the two-body spin-orbit interaction are dominant and quite sensitive to the
size of the model-space considered. The effects of the tensor interaction are
weaker. The correlations effects which are included in the larger (0+2)
$\hbar \omega$ shell-model space reduce the spin-orbit splitting in the case 
of $A=5$ by 20\%, and enhance it for $A=15$ by about the same 20\%. However, 
it is found that the correlations have a very small effect on the $d_{3/2} - 
d_{5/2}$ splitting in $A=17$.
\end{abstract}

\section{Introduction, Background and Motivation}

The problem of a microscopic understanding of the spin-orbit splitting in the
spectra of nuclei and its relation to the nucleon-nucleon interaction has
received a lot of attention over many years. As an example we quote from Bohr
and Mottelson \cite{bm} ``Finally, the tensor force contributes in
second-order and higher order to the effective one-body spin-orbit
potential''. They refer to the 1960 work of Terasawa $et~ al.$
\cite{ter,arima}.
Indeed, using the tensor interaction, these authors obtained a large 
spin-orbit splitting with the correct sign and level ordering. 
However, Terasawa noted that other groups got very small effects, some even 
of the opposite sign. One of the main motivations for Terasawa's work 
\cite{ter} was his feeling at that time ($i.e$ 1960) that it was not clear to 
what extent a two-body spin-orbit force would be required to explain the 
nucleon-nucleon ($NN$) data. 

The contributions to the spin-orbit splitting in the single-particle energies
for closed shell nuclei, which arise within the framework of a
non-relativistic solution of the nuclear many-body problem based on two-nucleon
interactions, have been investigated by Scheerbaum\cite{sch1,sch2}. His
investigations utilized an effective interaction\cite{sprung} which
corresponds to a parameterization of the Brueckner $G$-matrix  derived from
realistic interactions like the Reid soft-core potential\cite{reid}. Scheerbaum
demonstrated that a large part of the spin-orbit splitting can be attributed to
the effective $NN$ spin-orbit interaction contained in the Brueckner $G$
-matrix. 
This contribution occurs already in the mean field or Brueckner-Hartree-Fock
(BHF) approach.

In his work, Scheerbaum found that another important contribution to the 
spin-orbit splitting was related to the tensor
component in the $G$-matrix. This tensor force does not contribute to the
spin-orbit splitting of spin-saturated nuclei within the mean-field
approximation. A strong tensor force, however, leads to a sizeable contribution
of second order in this effective interaction. Since all particle-particle
ladder diagrams are already included in the Brueckner $G$-matrix, Scheerbaum 
only considered terms of second order in $G$ with intermediate hole-hole states. He observed 
that these terms of second order in $G$ lead to a contribution to the
spin-orbit splitting which is almost as large as  the effect due to the
spin-orbit component in the effective two-nucleon interaction\cite{sch1}.    

However, most of the so-called realistic $NN$ interactions, which were considered
around 1970, for instance the Reid soft-core potential\cite{reid}, contain a 
rather strong tensor component originating from the one-pion-exchange
contribution. On the other hand, One-Boson-Exchange models for the $NN$ 
interaction that have been developed more recently\cite{rupr} take into 
account the fact that this tensor component, originating from the 
one-pion exchange, is compensated to some extent by the contribution of the 
$\rho$-meson exchange, which yields a tensor contribution with an 
opposite sign. Therefore modern $NN$ interactions contain a weaker tensor
component than did previous ones like the Reid potential. This is one 
motivation to reanalyze the contribution to the spin-orbit splitting of the 
various components of the $NN$ interaction using a modern model of the $NN$ 
interaction.

A second motivation for our studies in the present paper is to investigate 
in a consistent and non-perturbative way the effect on the spin-orbit 
splitting of long-range correlations induced by the tensor force and other 
components of the $NN$ interaction. The short-range correlations leading to 
configurations with high-lying single-particle states are efficiently taken 
into account by means of the Brueckner $G$-matrix. However, the
effects of long-range correlations, involving shell-model configurations with
lower excitation energy, may require an explicit treatment. Therefore
one often splits the Hilbert space of all shell-model configurations into a
model-space (which includes, using the terminology of harmonic oscillator
states, all configurations up to $n\hbar \omega$) and the rest of the Hilbert
space. Correlations related to configurations outside the model space are
treated by determining an effective hamiltonian, which can be diagonalized
within the model space. 

A first approximation for this effective hamiltonian is to consider the
Brueckner $G$-matrix calculated from a solution of the Bethe-Goldstone equation
with a Pauli operator adjusted to the model space. Such effective hamiltonians,
based on the Bonn A and Bonn C\cite{rupr} potentials have been considered by 
Zamick et al.\cite{zzm2}. They investigated the single-particle energies using 
the BHF approximation supplemented by an explicit treatment of 2 particle - 1 
hole and 3 particle - 2 hole diagrams for configurations inside the model 
space. Inspecting the spin-orbit splitting, those authors found a strong 
cancellation between the 2-hole diagrams and the 2-particle diagrams. The 
effect of the hole-hole terms, which in agreement with the findings of 
Scheerbaum\cite{sch1} enhanced the spin-orbit splitting, was essentially 
compensated for by the corresponding particle-particle terms. 

The question is, does this cancellation hold beyond second-order perturbation
theory? To answer this question we are going to diagonalize the effective
hamiltonian in model spaces including configurations beyond one major shell. 
Furthermore we want to study the relative importance of the various terms in 
the effective $NN$ interaction, the central, tensor and spin-orbit parts of the 
NN interaction, as they contribute to the spin-orbit splitting. 
For that purpose we shall use a parameterization of the model-space $G$-matrix 
derived by Zheng and Zamick\cite{zheng}, which has the form

\begin{equation}
V(r)=V_c(r)+x \cdot V_{s.o.}+y \cdot V_t
\label{eq:xyint}
\end{equation}

\noindent where $s.o.$ stands for the two-body spin-orbit interaction, $t$
for the tensor interaction, and $V_c(r)$ is everything else, especially
the (spin-dependent) central interaction. The interaction terms $V_c$,
$V_{s.o.}$ and $V_t$ have been adjusted so as to obtain a good fit to the 
$G$-matrix elements for the Bonn A potential with $x=1,~y=1$. We can
study the effects of the spin-orbit and tensor interactions by varying
$x$ and $y$. More details about the interaction are given in reference
\cite{zheng}. 

The diagonalization of the effective hamiltonian in large model spaces is 
achieved by employing the OXBASH program\cite{oxbash}, taking care of spurious 
states by using the Gloeckner-Lawson technique\cite{gloe}.

In our present paper, we investigate the spin-orbit splitting within the
non-relativistic many-body approach using realistic two-body $NN$ interactions.
It has been demonstrated by Pieper et al.\cite{piep}, that three-nucleon 
forces may enhance the spin-orbit splitting, bringing it close to the 
experimental value in the case of $p_{1/2}$ and $p_{3/2}$ hole states in 
$A=15$. In fact they obtained a contribution of 2.84 MeV to this spin-orbit 
splitting from the Urbana VII model for the two-pion exchange three nucleon 
force\cite{urba} 

Another mechanism, which may be very important to the spin-orbit splitting, is
the change of the Dirac spinors for the nucleons in the nuclear medium as
predicted by the Dirac-Brueckner-Hartree-Fock approach\cite{broc,fritz}. The
strong and attractive scalar component in the relativistic self-energy for the
nucleon leads to an enhancement of the small components for the Dirac spinor in
the nuclear medium, which may be characterized in terms of a reduced Dirac mass
for the nucleons. This leads to an enhancement by about 2 MeV of the 
spin-orbit splitting of the $p_{1/2}$ and $p_{3/2}$ hole states in $A=15$
\cite{zzm2}. 

\section{Results}
\subsection{(a) The $A=5$ System}

We present in Table I the results of our 
shell-model calculations for the spin-orbit splitting $\Delta E=E({1/2}^-)-
E({3/2}^-)$ in mass A=5, considering three different model spaces: using a 
harmonic oscillator notation these model spaces are characterized by the 
excitation energies of the shell-model configurations that are 
included and denoted as 0 $\hbar \omega$, (0+2) $\hbar \omega$ and (0+2+4) 
$\hbar \omega$, respectively. Thus, 0$\hbar \omega$ corresponds to the case 
where we have a closed $0s$ shell and the one valence particle is in 
$0p_{3/2}$ or $0p_{1/2}$. For (0+2) $\hbar \omega$ we have the above valence 
configuration plus all 2 $\hbar \omega$ excitations, etc. In our case, due to 
computational limitations, the (0+2+4) $\hbar \omega$ space includes only the 
$0s$, $0p$, $0s-1d$ and $0f-1p$ shells, and is thus not quite complete. This 
is alos true for the (0+2) $\hbar \omega$ space in $^{17}O$. 

For the parameterization of the realistic $G$-matrix ($x=1,~y=1$ in
Eq. 1) we find that the values for the spin-orbit splitting
$\Delta E$ decrease with increasing sizes of the model-spaces. 
The values listed in Table I are  3.375 MeV, 2.959 MeV 
and 2.659 MeV in the 0, (0+2) and (0+2+4)
$\hbar \omega$ spaces, respectively.  
Thus, in higher order, we get a noticeable $reduction$ of
the effective spin-orbit splitting for $A=5$. What is the cause of
this reduction? Is it the two-body tensor interaction in play or 
the two-body spin-orbit interaction? To answer this question we performed
shell-model calculations varying the strength of the two-body spin orbit and 
the tensor interaction in terms of the variables $x$ and $y$ as defined in 
Eq. (1) and again show our results in Table I.

For $x=0,~y=0$, there is $no$ effective `spin-orbit' splitting ($ESO$)
in any of the model spaces. This reflects the fact that a central 
interaction, indeed even a {\em{spin-dependent}} central interaction, cannot 
induce any $ESO$ even if correlations in large model-spaces are considered. 

We also note that, in the 0 $\hbar \omega$ space, the $ESO$ is zero when 
$x=0$. The tensor interaction does not contribute to the $ESO$ for a spin 
saturated 
system ($i.e.$ for a closed $LS$ shell plus or minus one particle) if the 
mean-field approximation is considered (which corresponds to the calculation 
in the 0 $\hbar \omega$ space). As we vary $y$ (keeping $x=0$), we see an 
approximate quadratic rise in the effective spin-orbit splitting in each of 
the larger model spaces. In fact, the rise is a bit 
faster than quadratic in $y$. This shows that the tensor force contributes to
the $ESO$ only in second order and higher order of a perturbation expansion, 
with the dominant terms being of the form $V_t \times V_t$ in the notation of 
Eq. (1). The contribution of the tensor force to the $ESO$ has the correct 
sign.
However, that contribution is rather small for all reasonable values of $y$ 
($i.e.$ for $y \le 1$), and increases only by 5-9\% for any one of our values 
of $y$ in Table I when we include configurations of 4 $\hbar\omega$.

In Table I we also study in the $A=5$ system the effects of varying the 
two-body spin-orbit strength $x$ in the absence of the tensor interaction 
($i.e.$ for $y=0$). In the 0 $\hbar \omega$ space, the $ESO$ varies linearly
with $x$. We see the linear relation between the $ESO$ and
the two-body spin-orbit interaction in the mean-field approach. Interestingly,
also in the larger spaces the $ESO$ varies almost linearly with $x$. This 
indicates that a 
perturbative inclusion of these correlations in the larger model space would
be dominated by terms of the form $V_c \times V_{s.o.}$ using the nomenclature
of Eq. (1).

Perhaps the most important result of Table I is that for $A=5$, there is a
systematic $decrease$ in the spin-orbit splitting as one goes to
larger model spaces. For example, in the 0, (0+2) and (0+2+4) $\hbar \omega$
spaces the values of the $ESO$ for $x=1$ ($y=0$) are 3.375, 2.716 
and 2.431 $MeV$, respectively. In each of the three cases that 
we studied in Table I: $(x,y)=(0.5,0),~(1,0),~(1.5,0)$, we find that going 
from 0 $\hbar 
\omega$ to (0+2) $\hbar \omega$ decreases the $ESO$ by about 20\%, and going 
from (0+2) $\hbar \omega$ to (0+2+4) $\hbar \omega$ in each case further 
decreases the $ESO$ by about 10\%. The percentages of change are the same in 
all three cases due to the linearity of the $ESO$'s with $x$. 

While there has been some discussion of
the enhancement of the spin-orbit interaction for $A=5$ due to
second-order tensor effects \cite{ter}, we are not aware of any
discussion of the spin-orbit interaction in higher order.

We see the combined effects of the spin-orbit and tensor interactions
by comparing the $x=1,~y=1$ case in Table I with the $x=1,~y=0$ case. The 
small effects of the 
tensor force can essentially be added to the results obtained with the central
plus spin-orbit interaction. This is true both in the (0+2) $\hbar \omega$ and 
in the (0+2+4) $\hbar \omega$ spaces. In each case, the contribution of the 
tensor interaction is less than 10\% of that of the spin-orbit interaction. 

The observed values for the ${1/2}^- - {3/2}^-$ level separation in the 
$A=5$ system have large experimental uncertainties, being $7.5 \pm 2.5~MeV$ 
for $^5Li$ and $4.0 \pm 1.0~MeV$ for $^5He$ \cite{ndata}. The calculated 
results in Table I agree with the observed data better for $x=1.5$ than for 
$x=1.0$. Such an enhancement of the strength of the two-body spin-orbit 
interaction in actual nuclei was also suggested for nuclei in the beginning of 
the $1s-0d$ shell by the work of Fayache $et~ al.$ \cite{npa}.

\subsection{The $A=15$ System}

Next we consider the $E({3/2}_1^-)-E({1/2}_1^-)$ splitting in mass 15. In
lowest order (0 $\hbar\omega$) these states are described as hole states
relative to the closed shell nucleus $^{16}O$. In that picture, the ground 
state of the $A=15$ system is a $p_{1/2}$ hole, and the first excited state
is a $p_{3/2}$ hole. The results for the calculations in the 0 $\hbar \omega$ 
and (0+2) $\hbar \omega$ spaces are presented in Table II.

In the $A=15$ system, and for $y=0$ (no tensor interaction), the $ESO$ is 
linear in $x$ in the 0 $\hbar \omega$ space and very nearly linear in $x$ in 
the (0+2) $\hbar \omega$ space. For the $x \neq 0,~y=0$ cases, for $each$ $x$ 
value ($x=0.5,~1.0$ or 1.5), the $ESO$ for $A=15$ $increases$ by about 20\% 
as we go from 0 $\hbar \omega$ to (0+2) $\hbar \omega$. We recall that under 
these circumstances, the $ESO$ for the $A=5$ system $decreased$ by about 20\%. 
We again understand in the $A=15$ system that the percentages of change for 
the three $x$ values ($x \neq 0,~y=0$) are the same because of the linearity 
of the $ESO$'s with $x$ (for $y=0$) in both of the spaces considered. However, 
it is interesting to note that (but harder to explain why) there is an 
increase in $A=15$ but a decrease in $A=5$, and also why the percent change 
in both systems is the same (about 20\%). The linearity with $x$ (for $y=0$) 
indicates again that the corrections to the $ESO$'s in second order are 
dominated by terms of the form $V_c \times V_{s.o.}$.

For the $A=15$ system, when we vary the tensor interaction with the 
spin-orbit interaction turned off ($x=0$), we get again a nearly quadratic 
dependence of the $ESO$ on $y$. Once more, the magnitude of the $ESO$ rises 
slightly faster than quadratically with $y$. This shows that, for the 
$A=15$ system as well, the tensor force contributes to the $ESO$ only in 
second and higher orders of a perturbation expansion with the dominant terms 
being of the form $V_t \times V_t$. The contribution of the tensor term by 
itself to the $ESO$ ($i.e.$ when $x=0$) is very small (an order of magnitude 
smaller than its already small contribution in the $A=5$ case), and has the
wrong sign. 

For the $A=15$ system, and in the 0 $\hbar \omega$ space where the tensor 
interaction $cannot$ contribute, the $ESO$ is 5.063 $MeV$ for both the $x=1,
~y=0$ and the $x=1,~y=1$ cases. In the (0+2) $\hbar \omega$ space the $ESO$ is
6.008 $MeV$ for $x=1,~y=0$ and 5.698 $MeV$ for $x=1,~y=1$. We thus see from 
Table II that the effect of the tensor interaction (with $y=1$) is more 
significant (-0.32 $MeV$) when the spin-orbit interaction is present ($x=1$) 
than when the spin-orbit interaction is absent (-0.009 $MeV$ for $x=0$).
For $A=15$, and unlike the $A=5$ case, the tensor and the spin-orbit effects 
are not additive, indicating the presence in $A=15$ (but not in $A=5$) of a 
larger second-order term of the form $V_{s.o.} \times V_t$.

When for $A=15$ we consider the realistic interaction $x=y=1$, we see again 
that in both spaces the $ESO$ is very largely due to the spin-orbit 
interaction, while the effect of the tensor interaction is small and of the 
wrong sign. In contrast to the $A=5$ case, the $ESO$ in the $A=15$ system 
for $x=1,~y=1$ is larger in the (0+2) $\hbar \omega$ space than in the 
0 $\hbar \omega$ space, with most of the enhancement again due to the 
spin-orbit interaction. 

For $A=15$, the observed $E({3/2}_1^-)-E({1/2}_1^-)$ splitting is 6.324 $MeV$ 
for $^{15}N$ and 6.176 $MeV$ for $^{15}O$ \cite{ndata}. The results of Table 
II suggest that for $A=15$, including the 2 $\hbar \omega$ excitations 
and taking $x=1$ 
lead to results in closer agreement with the observed level separations.

\subsection{The A=17 System}

The results of calculations of the $ESO$ for $A=17$, considering  the
spin-orbit partners are $0d_{5/2}$ and $0d_{3/2}$, are given in 
Table III. 
For the realistic $x=1,~y=1$ interaction, there is hardly any change (a mere 
increase of 0.1 $MeV$) in the $ESO$ in going from 0 $\hbar \omega$ to (0+2) 
$\hbar \omega$. Again, for $y=0$ the $ESO$'s are proportional to $x$ in the 
0 $\hbar \omega$ space and very nearly so in the (0+2) $\hbar \omega$ space. 
For all the $y=0$ cases, the effect on the $ESO$'s of going from 
0 $\hbar \omega$ to (0+2) $\hbar \omega$ is an increase, but by less than 
3\%. The small enhancement of the $ESO$ for $x=1,~y=1$ as we go to the larger 
space is again largely due to the two-body spin-orbit interaction. The effect 
of the tensor force is again very weak; for $x=0$ and a variable $y$, the 
$ESO$ is again zero in the 0 $\hbar \omega$ space and has a magnitude of 
0.02 $MeV$ or less for $y \le 2$ in the (0+2) $\hbar \omega$ space. For $x=0$ 
in the (0+2) $\hbar \omega$ space, the behavior of the $ESO$ as a function of 
$y$ is rather complicated and even changes sign. It starts from $y=0$ by being 
very slightly negative and reaches a minimum of about -0.01 $MeV$ for $y 
\approx 1$. The $ESO$ then increases as $y$ increases further, becoming 
positive for $y \ge 1.63$. Indeed, for $x=0$ and $y \ge 1.63$, $\Delta ESO$ 
shows a rapid but less than quadratic increase with $\Delta y$. These last  
observations can be taken as a possible indication that there is a 
cancellation between two effects. This would support the results of the 
calculations of \cite{zzm2} in which Zheng $et~al.$ observed that in 
perturbation theory there is a similar cancellation between the contributions 
from 2 particle-1 hole states and those from 3 particle-2 hole states. 

It is interesting to study the variation with $x$ and $y$ of two other 
quantities in the $A=17$ system. One is $E(1s_{1/2})$, the energy of the 
$1s_{1/2}$ level, and the other is $E_{com}$, the energy of the center of mass 
of the $0d_{5/2}-0d_{3/2}$ spin-orbit doublet, where 

\begin{equation}
E_{com} \equiv 0.6E(0d_{5/2})+0.4E(0d_{3/2})
\end{equation}

In the 0 $\hbar\omega$ space, both the $E(1s_{1/2})$ and the $E_{com}$ are 
strictly independent of both $x$ and $y$. In that small space (with one 
particle being outside a closed-shell core and no particle-hole excitations), 
the two-body tensor force has no effect on the $E(1s_{1/2})$, $E(0d_{5/2})$, 
or $E(0d_{3/2})$, and hence no effect on the $E_{com}$. This explains why 
in Table III, both the $ESO$ and $E(1s_{1/2})-E(0d_{5/2})$ are independent of 
$y$ in the 0 $\hbar\omega$ space. 

Furthermore, in the 0 $\hbar\omega$ space the two-body spin-orbit interaction 
acts like a one-body spin-orbit force, and thus it has no effect either on the 
$E_{com}$ or on the energy of the $1s_{1/2}$ level which has $l=0$. Hence, 
in the 0 $\hbar \omega$ space, both 
the $E_{com}$ and the $E(1s_{1/2})$ are unaffected also by changes in $x$. As 
$x$ increases from $x=0$, $E(0d_{5/2})$ decreases by an amount $\delta$,  
proportional to $x$, while $E(0d_{3/2})$ increases by 1.5$\delta$, so that 
indeed $E_{com}$ is left unchanged. 

In the 0 $\hbar\omega$ space we calculate that for all $x,~y$ values, 

\begin{equation}
E_{com} - E(1s_{1/2}) = 2.343 ~MeV
\end{equation}

\noindent This energy difference is due to the attractive central force 
component $V_c(r)$ in the effective $NN$ interaction of Eq. (1). The 
experimental data \cite{ndata} for $^{17}O$ shows a ${5/2}^+$ ground state 
with excitation energies (in $MeV$) of 0.871 for ${1/2}^+$ and 5.084 for 
${3/2}^+$. With this observed data, $E_{com} - E(1s_{1/2})$ is equal to 1.465 
$MeV$. From Eqs(2) and (3) and the definition of the $ESO$, we obtain in the 
0 $\hbar\omega$ space the following relationship: 

\begin{equation}
\left [ E(1s_{1/2}) - E(0d_{5/2})\right ] -0.4 ESO = -2.343 ~MeV
\end{equation}

All the 0 $\hbar\omega$ data in Table III is fitted perfectly by this 
relationship. In the (0+2) $\hbar\omega$ space, and for $x=0~y=0$, we have 
$E(0d_{5/2})=E(0d_{3/2}) \equiv E_{com}$, and due to the central force term 
in Eq. (1) we calculate that 

\begin{equation}
E_{com} - E(1s_{1/2}) = 3.853 ~MeV
\end{equation}

In the (0+2) $\hbar\omega$ space, and keeping $y=0$, we note that as $x$ 
increases a relationship similar to Eq. (4) but with -2.343 replaced by 
-3.853 holds to within 3\% or better (see Table III). In this large space, 
however, and keeping $x=0$, we note that an increase in $y$ also renormalizes 
the central force term. All three energies $E(0d_{5/2}),~E(0d_{3/2})$ and 
$E(1s_{1/2})$ decrease in the large space with increasing $y$. 
The $E(0d_{5/2})$ and the 
$E(0d_{3/2})$ (and hence the $E_{com}$) all decrease at the same rate. Hence, 
in Table III, and for (0+2) $\hbar\omega$, the $ESO$'s are very small for 
$x=0$ and any $y$. But the above rate of decrease is about 15\% larger than 
the corresponding rate of decrease for the $E(1s_{1/2})$. Hence, for 
$x=0$, as $y$ increases in Table III, the separation $E(1s_{1/2})-
E(0d_{5/2})$, which is calculated to be negative for $y=0$ in the (0+2) 
$\hbar\omega$ space, becomes less negative with increasing $y$. 

\section{Additional Remarks and Summary}

To summarize our paper, we have shown that the contribution to the effective 
spin-orbit splitting of the two-body tensor term in the effective interaction 
is generally much smaller than the contribution from the two-body spin-orbit 
term. This result is a consequence of the weaker tensor component in modern 
$NN$ interactions. An earlier investigation \cite{sch1,sch2} used older 
models of realistic interactions with stronger tensor components and obtained 
a much larger contribution from the tensor term to the effective spin-orbit 
splitting. 

We see in our non-perturbative calculations that the effects of higher-shell 
admixtures on the $ESO$'s cannot be ignored, being generally in about the 10-
20\% range for $A=5$ and $A=15$, but less than 3\% for $A=17$. 

Additional insights into some of our results are provided by the perturbative 
work of reference \cite{zzm2}. A comparison with their results suggests 
similarities between the behavior of the second-order two-body spin-orbit 
interaction term and some Hartree-Fock type diagrams. For both the $A=5$ and 
$A=15$ systems, there do not seem to be nearly complete cancellations of the 
effects of particle-particle admixtures and hole-hole admixtures, which 
respectively tend to reduce and enhance the $ESO$ \cite{zzm2} 
(see also \cite{sch1}). 
In the $0p$-shell nuclei, the reduction effects are more important for $A=5$ 
(one particle beyond a closed shell), while the enhancement effects prevail 
for $A=15$ (one hole away from a closed shell). However, for $A=17$ the 
cancellation is nearly complete, and going from 0 $\hbar \omega$ to 
(0+2) $\hbar \omega$ increases the $ESO$ by less than 3\%. 

We found that even in the larger (0+2) $\hbar \omega$ space there is an 
almost linear relationship between the $ESO$ and the strength of the two-body 
spin-orbit component of the effective interaction. On the other hand, in this 
larger space, the $ESO$'s dependence on the strength of the two-body tensor 
component of the effective intreaction is close to being quadratic. 

Recalling the situation for $A=15$ and $A=17$, we find that the
effect of  2 $\hbar \omega$ configurations yields a larger
enhancement for $A=15$ than for $A=17$. The spin-orbit splittings
for the 
particle states tend to be reduced as compared to those for hole
states, as noted in ref.
\cite{zzm2}. This is supported from 
experiment. The splitting $E({3/2}_1^-)-E({1/2}_1^-)=6.0~MeV$  is 
larger than the corresponding $A=17$ splitting 
$E({3/2}_1^+)-E({5/2}_1^+)=5.1~MeV$ (although the orbital angular
momentum is larger in the latter case) and the corresponding 
$E({1/2}_1^-)-E({3/2}_1^-)= 4.0~ MeV $ splitting in $A=5$. Thus, 
large space calculations are 
essential in this context to properly account for the differences in
spin-orbit splittings of single particle states above the Fermi energy
and of single hole states below the Fermi energy. 

Finally, the correlation effects which are included in the (0+2) 
$\hbar \omega$ shell model space $reduce$ the contribution to the $ESO$ of 
the dominant two-body spin-orbit interaction term by about 20\% in the $A=5$ 
system, but $increase$ the contribution by about the same 20\% in the $A=15$ 
system. The fact that both magnitudes are essentially the same requires 
further investigation. 

\section{Acknowledgements:}
Support from the U.S. Department of energy DE-FG 02-95ER-40940 is greatly 
appreciated. Y.Y. Sharon gratefully ackowledges a Summer 2000 Richard Stockton 
College Distinguished Faculty Fellowship.

\samepage

\begin{table}
\caption{The Effective Spin-Orbit splitting $ESO=E({1/2}^-)-E({3/2}^-)$
for $A=5$ varying $y$ (strength of tensor force) and $x$ (strength of the 
two-body spin-orbit interaction).}

\begin{tabular}{rr|rrr}
$x$ & $y$ & \multicolumn{3}{c}{$ESO$ [$MeV$]}\\
\tableline
 & & 0 $\hbar \omega$ & (0+2) $\hbar \omega$ & (0+2+4) $\hbar
\omega$\\ 
\tableline
1 & 1   & 3.375 & 2.959 & 2.659\\
\tableline
0 & 0 & 0 & 0 & 0\\
0 & 0.5 & 0 & 0.046 & 0.050\\
0 & 1   & 0 & 0.216 & 0.230\\
0 & 1.5 & 0 & 0.542 & 0.572\\
0 & 2   & 0 & 1.034 & 1.092\\
0 & 3   & 0 & 2.457 & 2.640\\
\tableline
0.5 & 0 & 1.688 & 1.375 & 1.238\\
1   & 0 & 3.375 & 2.716 & 2.431\\
1.5 & 0 & 5.063 & 4.012 & 3.584\\
\end{tabular}
\end{table}

\begin{table}
\caption{The ${3/2}^- - {1/2}^-$ splitting 
in $A=15$ with various $x$ and $y$ combinations.} 
\begin{tabular}{cc|cc}
$x$ & $y$ & \multicolumn{2}{c}{$ESO$ [$MeV$]}\\
\tableline
 & & 0 $\hbar \omega$ & (0+2) $\hbar \omega$\\
\tableline
1   & 1 & 5.063 & 5.698\\
\tableline
0   & 0   &   0 &   0 \\
0   & 0.5 &   0 & -0.002\\
0   & 1   &   0 & -0.009\\
0   & 1.5 &   0 & -0.019\\
0   & 2   &   0 & -0.036\\
0   & 2.5 &   0 & -0.059\\
0   & 3   &   0 & -0.088\\
\tableline
0.5 & 0 & 2.531 & 3.026\\
1   & 0 & 5.063 & 6.008\\
1.5 & 0 & 7.593 & 8.934\\
\end{tabular}
\end{table}

\begin{table}
\caption{The ${3/2}^+ - {5/2}^+$ splitting 
in $A=17$, as well as the $1s_{1/2}$ energy (relative to $0d_{5/2}$
for various $x$ and $y$ combinations)$^a$.}  
\begin{tabular}{cc|cc|cc}
$x$ & $y$ & { 0 $\hbar \omega$} &
{(0+2) $\hbar \omega$}& { 0 $\hbar \omega$} &{(0+2) $\hbar \omega$}\\
\tableline
$x$ & $y$ & \multicolumn{2}{c}{$ESO$ [$MeV$]} &\multicolumn{2}{c}
{$E(1s_{1/2})-E(0d_{5/2})$ [$MeV$]} \\
\tableline
1   & 1 & 5.562  & 5.662 & -0.119& -1.430\\
\tableline
0   & 0   & 0 & 0 &-2.343 &  -3.853\\
0   & 0.5 & 0 & -0.005 & -2.343 & -3.806\\
0   & 1   & 0 & -0.010 & -2.343 & -3.661\\
0   & 1.5 & 0 & -0.004 & -2.343 & -3.419\\
0   & 2   & 0 & 0.024 & -2.343 &  -3.085\\
0   & 2.5   & 0 & 0.078 & -2.343 &  -2.671\\
0   & 3   & 0 & 0.160 & -2.343 &  -2.195\\
\tableline
0.5 & 0 & 2.782 &2.849 & -1.231 &  -2.723\\
1   & 0 & 5.562 & 5.689 & -0.119 & -1.618\\
1.5 & 0 & 8.344 & 8.522 & 0.994 & -0.534\\
\end{tabular}
(a) In all cases, at the 0 $\hbar \omega$ level, $E_{com}-E(1s_{1/2})=2.343~
MeV$. For the (0+2) $\hbar \omega$ case, there is at most a 3\% deviation 
from the $x=0,~y=0$ value of 3.853 $MeV$.
\end{table}


\begin{references}
\bibitem{bm} A. Bohr and B. Mottelson, {\em Nuclear Structure}, 
W.A. Benjamin Inc., New York, 1975, Vol. I, p. 259.

\bibitem{ter} T. Terasawa, Prog. Theor. Phys. (Kyoto), {\bf 23},
87 (1960). 

\bibitem{arima} A. Arima and T. Terasawa, Prog. Theor. Phys. (Kyoto),
{\bf 23}, 115 (1960).

\bibitem{sch1} R.R. Scheerbaum, Phys. Lett. {\bf B 61}, 151 (1976).

\bibitem{sch2} R.R. Scheerbaum, Nucl. Phys. {\bf A 257}, 77 (1976).

\bibitem{sprung} D.W.L. Sprung, Nucl. Phys. {\bf A 182}, 97 (1972).

\bibitem{reid} R.V Reid, Ann. Phys. (NY) {\bf 50}, 411 (1968). 

\bibitem{rupr} R. Machleidt, Adv. Nucl. Phys. {\bf 19}, 189 (1989).

\bibitem{zzm2} D.C. Zheng, L. Zamick and H. M\"uther,
        Phys. Rev. {\bf C 45}, 2763(1992).

\bibitem{zheng} D.C. Zheng and L. Zamick, Ann. Phys. (NY) {\bf 206}
106(1991).

\bibitem{oxbash} B. A. Brown, A. Etchegoyen and W. D. M. Rae, The
computer code OXBASH, MSU-NSCL report number 524, 1992.

\bibitem{gloe} D.H. Gloeckner and R.D. Lawson, Phys. Lett. {\bf
B 53}, 313(1974).

\bibitem{piep} S.C. Pieper and V.R. Pandharipande, Phys. Rev. Lett. {\bf 70},
2541 (1993).

\bibitem{urba} R. Schiavilla, V.R. Pandharipande, and R.B. Wiringa, Nucl. Phys.
{\bf A449}, 219 (1986).

\bibitem{broc} H. M\"uther, R. Machleidt, and R. Brockmann, Phys. Rev. {\bf C 
42}, 1981 (1990).

\bibitem{fritz} R. Fritz, H. M\"uther, and R. Machleidt, Phys. Rev. Lett. {\bf
71}, 46 (1993).

\bibitem{ndata} Nuclear Data Retrieval ($http://www.nndc.bnl.gov$).

\bibitem{npa} M. S. Fayache, P. von Neumann-Cosel, A. Richter, Y. Y. Sharon 
and L. Zamick, Nucl. Phys. {\bf A 627}, 14 (1997).
\end{references}
\end{document}